\documentclass[11pt,aps,nofootinbib,floatfix,preprint]{revtex4}
\usepackage{mathrsfs}
\usepackage{graphicx}
\usepackage{amsmath}
\usepackage{amsfonts}
\usepackage{amssymb}
\usepackage{color}

\usepackage{epsfig}
\usepackage{CJK}
\usepackage{graphicx}
\usepackage{epsfig}
\usepackage{eepic}
\usepackage{bbm}
\usepackage{dcolumn}
\usepackage{bm}
\usepackage{ulem}
\usepackage{slashbox}
\usepackage{multirow}

\newcommand{\omits}[1]{}

\def\bc{\begin{center}}
\def\nno{\nonumber}
\def\ec{\end{center}}
\def\be{\begin{eqnarray}}
\def\ee{\end{eqnarray}}

\definecolor{dyellow}{rgb}{1.,0.8,.0}
\definecolor{myblue}{rgb}{.1,.1,.7}
\definecolor{dcyan}{rgb}{.0,.6,.6}
\definecolor{cyan}{rgb}{0.4,1.0,1.0}
\definecolor{dmagenta}{rgb}{0.6,0.0,0.6}
\definecolor{brown}{rgb}{0.6,0.2,0.}
\definecolor{darkblue}{rgb}{.0,.0,0.5}
\definecolor{darkred}{rgb}{0.75,0.0,0.0}
\definecolor{orange}{rgb}{1.,.6,.0}
\definecolor{dorange}{rgb}{0.8,.4,.0}
\definecolor{green}{rgb}{0.0,1.0,0.0}
\definecolor{darkgreen}{rgb}{0.0,0.6,0.0}
\definecolor{purple}{rgb}{.4,.0,.4}
\definecolor{lightgrey}{rgb}{0.7, 0.7, 0.7}
\definecolor{grey}{rgb}{0.4, 0.4, 0.4}

\def\La{\Lambda}

\def\al{\alpha}
\def\ga{\gamma}

\def\la{\lambda}

\def\si{\sigma}
\def\om{\omega}

\newcommand{\nc}{\newcommand}
\nc{\rnc}{\renewcommand} \nc{\ket}[1]{\left | \, #1 \right \rangle}
\nc{\bra}[1]{\left \langle #1 \, \right |}
\nc{\ua}{\uparrow} \nc{\da}{\downarrow}

\nc{\braket}[2]{\langle\, #1\,|\,#2\,\rangle}
\nc{\half}{\frac{1}{2}}

\nc{\prj}{\mathcal{P}} \nc{\hilb}{\mathcal{H}}
\nc{\pth}{\mathcal{C}} \nc{\inprod}[2]{\braket{#1}{#2}}
\nc{\upket}{\ket{\uparrow}} \nc{\downket}{\ket{\downarrow}}
\nc{\upbra}{\bra{\uparrow}} \nc{\downbra}{\bra{\downarrow}}

\begin{document}


\title{Mimic the optical conductivity in disordered solids via gauge/gravity duality}

\author{Jia-Rui Sun$^{1}$} \email{jrsun@ecust.edu.cn}
\author{Shang-Yu Wu$^2$} \email{loganwu@gmail.com}
\author{Hai-Qing Zhang$^3$} \email{hqzhang@cfif.ist.utl.pt}

\affiliation{${}^1$Department of Physics and Institute of Modern
Physics, East China University of Science and Technology, Shanghai
200237, China} \affiliation{${}^2$Institute of Physics,
National Chiao Tung University, Hsinchu 300, Taiwan\\
National Center for Theoretical Science, Hsinchu, Taiwan\\
Yau Shing Tung Center, National Chiao Tung University, Hsinchu, Taiwan}
\affiliation{${}^3$CFIF, Instituto Superior {T}\'ecnico,
Universidade {T}\'ecnica de Lisboa, Av. Rovisco Pais 1, 1049-001
Lisboa, Portugal}



\begin{abstract}
 We study the optical conductivity in a (2+1)-dimensional non-relativistic field theory holographically dual to a (3+1)-dimensional charged Lifshitz black brane within the Einstein-Maxwell-dilaton theory. Surprisingly, we find that the optical AC conductivity satisfies the nontrivial (non-)power law scaling in the high frequency regime rather than approaching to a constant when the dynamical critical exponent $z>1$, which is qualitatively similar to those in various disordered solids in condensed matter systems. Besides, this (non-)power law scaling behavior shows some universality, which is robust against the temperatures. We argue that the peculiar scaling behavior of AC conductivity may stem from the couplings of the dilaton field with the gauge fields and also the logarithmic behavior near the boundary in the Lifshitz spacetime.
\end{abstract}


\maketitle

\newpage


\section{Introduction}
The AdS/CFT correspondence \cite{Maldacena:1997re,Gubser:1998bc,Witten:1998qj} or generally the gauge/gravity duality, has provided us very effective tools to study the properties of the strongly coupled quantum field theories which live on the boundary of certain gravitational backgrounds. One of the most important characteristics of the gauge/gravity duality is that it is a kind of strong-weak duality. In view of this duality, various important phenomena of the strongly coupled field theories can be studied by performing calculations on their dual weakly coupled gravity side. Recently, motivated from the study in condensed matter physics, many attempts have been made in constructing bulk gravitational solutions to model numerous types of strongly coupled phenomena in condensed matter systems, especially close to the phase transition or quantum critical points, including the superconductor (superfluid) phase transition \cite{Hartnoll:2008vx}, Fermi and non-Fermi liquids \cite{Lee:2008xf}, superconductor-insulator transitions \cite{Nishioka:2009zj} etc, for recent review, see \cite{Hartnoll:2009sz,Iqbal:2011ae}. There are also some quantum phase transition systems in condensed matter physics which contain the Lifshitz-fixed points have received much attention. On one hand, these developments successfully extended the gauge/gravity duality into the more general form, namely, non-relativistic version \cite{Son:2008ye,Balasubramanian:2008dm,Kachru:2008yh,Kovtun:2008qy,Taylor:2008tg,Danielsson:2009gi,Alishahiha:2012nm,Mozaffar:2012bp}. On the other hand, they allow us to study strongly coupled systems toward realistic laboratory conditions by holographic principle, which may also be used as a test for the gauge/gravity duality itself.

This paper focuses on dealing with Lifshitz field theory in the framework of the non-relativistic gauge/gravity duality. Following our previous work \cite{Sun:2013wpa}, we continue studying the holographic optical conductivity in the quantum field theory which is dual to the Lifshitz black brane with two independent $U(1)$ gauge fields \cite{Tarrio:2011de}. To compare with the phenomena in condensed matter physics, we work in a $(3+1)$-dimensional Lifshitz spacetime, {\it i.e.}, the dual field theory is $(2+1)$-dimensional, and we focus on the case of $1\leq z \leq 2$, where $z$ is the dynamical critical exponent. When $z=1$, the Lifshtiz black brane will return to the usual Reissner-Nordstr\"{o}m (RN)-AdS black brane, therefore, the optical conductivity we obtain is similar to those studied in previous AdS/condensed matter literatures, such as \cite{Hartnoll:2009sz}. However, when $z>1$ the optical conductivity, especially its AC part, shows interesting behavior which is less discussed before as far as we know. More explicitly, we find that the optical conductivity will possess a non-trivial scaling with respect to the frequency in large frequency regime when $z>1$, such as $\om^{s(z)}$ where $s(z)>0$ is a function of $z$. This feature is very interesting, since in the previous literatures people argued that the large frequency behavior of the electric conductivity in $(2+1)$-dimensional field theory will approach to a constant due to the dimensionless of the conductivity \cite{Hartnoll:2009ns}. While in this paper we do find a counterexample. We argue from the viewpoint of non-relativistic gauge/gravity duality that the particular scaling behavior of the optical conductivity with respect to large frequency is caused by the couplings between the dilaton and the electromagnetic fields in the Einstein-Maxwell-dilaton (EMD) theory. This kind of non-minimal coupling will probably introduce extra dimensional scale into the boundary field system, which results in the peculiar evolution of the optical  conductivity associated with the frequency. For earlier studies on holographic properties of charged dilatonic black brane in EMD theory with non-minimal coupling in asymptotically AdS spacetime, see for example \cite{Cadoni:2009xm,Cadoni:2011kv}.

More interestingly, we surprisingly find from the condensed matter literatures that there indeed exists a similar universal behavior of the optical conductivity in (2+1)-dimensional condensed matter systems, such as \cite{Dyre:2000zz}, in which the authors studied the optical conductivity in various disordered solids both experimentally and theoretically. We observe that the optical conductivity studied in the present holographic model has very like behaviors to those in disordered solids in the extreme disorder limit for both high and low frequencies, at least qualitatively. In order to figure out this interesting phenomenon in more detail, we extend our previous study on the optical conductivity \cite{Sun:2013wpa} into various temperatures and find that in the low frequency regime, the optical conductivity will decrease as the temperature decreases, which is consistent with the experiments in \cite{Dyre:2000zz} qualitatively.  In particular, at zero temperature the conductivity will vanish, which suggests that the conducting electrons will be frozen at zero temperature \footnote{This does not conflict with the fact that the extremal Lifshitz black brane has a nonvanishing entropy, since according to the black hole/CFT correspondence, the entropy of extremal black hole is contributed from the ground state degeneracy of the near horizon microstates. It would also be interesting to study the microscopic entropy of this Lifshitz black brane by extending the methods in the RN/CFT correspondence \cite{Chen:2009ht,Chen:2010as,Chen:2010yu,Chen:2010ywa,Chen:2011gz}. However, this is beyond the scope of the present paper.}. In addition, we show that all the optical conductivities will have the same scaling with respect to the frequency whatever the temperature is for the fixed $z$ in the high frequency regime. This robust phenomenon for the high frequency behavior of the conductivity is also in accordance with the experiments \cite{Dyre:2000zz}. Furthermore, we also find a linear relation between the logarithmic of the optical conductivity versus the reciprocal of the temperature in some regime of the temperatures, which is qualitatively similar to those in disordered solids \cite{Psarras:2006}, as well. All of these consistencies of the holographic optical conductivity with those in various disordered solids allow us to guess that there might be some deep relationship between them, although the underlying precise details are not very clear at present. Strictly speaking, there are no apparent disorder parameters in our model, namely, there is neither spatial inhomogeneity in the background spacetime, nor interaction terms randomly distributed on the spatial coordinates like those studied in \cite{Hartnoll:2008hs,Fujita:2008rs,Ryu:2011vq,Hashimoto:2012pb}. However, we want to point out that the fluctuation of the second $U(1)$ gauge field in our construction could be interpreted as the impurity field \cite{Park:2013goa}, which interacts with the first $U(1)$ gauge field indirectly through the dilaton field. The homogeneous optical conductivity may relate the extreme disorder limit of disordered solids in which local randomly varying mobilities of charge carriers cover many orders of typical length scale of the condensed matter system \cite{Dyre:2000zz}.

The rest parts of this paper is organized as follows: The configuration of the asymptotic Lifshitz brane is briefly introduced in Section \ref{sect:intro}; We show the numerical results of the optical conductivity in Section \ref{sect:conduc}; The conclusions and discussions are drawn in Section \ref{sect:conclusion}.

\section{The Configuration of the Lifshitz Black Brane }\label{sect:intro}
The bulk gravitational theory we consider is the $(3+1)$-dimensional Einstein-Maxwell-dilaton (EMD) theory with the action
\begin{equation}\label{EMd1}
I=\frac{1}{16\pi G_4}\int{d^4x}\sqrt{-g}\left(R-2\La-\frac{1}{4}e^{\la_1\phi}F_1^{2}-\frac{1}{4}e^{\la_2\phi}F_2^{2}-\frac{1}{2}(\partial\phi)^2\right),
\end{equation}
where $\Lambda$ is the cosmological constant, $F_{a\mu\nu}=\partial_\mu A_{a\nu}-\partial_\nu A_{a\mu}$ $(a=1,2)$ are the $U(1)$ gauge field strengths associated with two independent gauge fields $A_{1\mu}$ and $A_{2\mu}$, $\phi$ is the dilaton field, while $\la_1$ and $\la_2$ are the coupling constants between the gauge fields and the dilaton. The dynamical equations in the bulk are
\begin{eqnarray}
\Box\phi &=&\frac{1}{4}\sum^2_{a=1}\la_a e^{\la_a\phi} F^{2}_a, \nonumber\\
\nabla_{\mu}(e^{\la_a\phi}F_a^{\mu\nu})&=& 0, \nonumber\\
R_{\mu\nu}-\frac{1}{2}g_{\mu\nu}R &=& \frac{1}{2}\sum^2_{a=1}e^{\la_a\phi}\left(F_{a\mu\la}F_{a\nu}
^{\la}-\frac{1}{4}g_{\mu\nu}F_a^2\right)+\frac{1}{2}\left(\partial_{\mu}\phi\partial_{\nu}\phi-\frac{1}{2}g_{\mu\nu}(\partial\phi)^{2}\right).
\end{eqnarray}
One of the solutions for the above EMD theory is a kind of charged Lifshitz black brane derived in \cite{Tarrio:2011de}
\begin{eqnarray}\label{bgd}
&&ds^{2}=-\frac{r^{2z}}{l^{2z}}f(r)dt^{2}+\frac{l^{2}}{r^{2}f(r)}dr^{2}+\frac{r^{2}}{l^2}\left(dx^2+dy^2\right), \nonumber\\
&&f(r)=1-\frac{m}{r^{z+2}}+\frac{\mu^{-\sqrt{z-1}}l^{2z}q^2}{4z r^{2(z+1)}}, \nonumber\\
&&A'_{1t}=l^{-z}\sqrt{2(z+2)(z-1)}\mu^{\sqrt{\frac{1}{(z-1)}}}r^{z+1},\quad A'_{2t}=\frac{q\mu^{-\sqrt{z-1}}}{r^{z+1}},\nonumber\\
&&e^{\phi}=\mu r^{\sqrt{4(z-1)}}, \quad\lambda_1=-\sqrt{\frac{4}{z-1}},\quad\lambda_2=\sqrt{z-1},\quad \Lambda=-\frac{(z+2)(z+1)}{2l^2},
\end{eqnarray}
where $l$ is the curvature radius of the Lifshitz spacetime, $\mu$ is the scalar field amplitude, $m$ and $q$ are respectively related to the mass and charge of the black brane.  The Hawking temperature of the Lifshitz black brane and the entropy density of boundary non-relativistic field are respectively
\be
T=\frac{1}{4\pi}\left(\frac{r_h}{l}\right)^{z+1}\left(\frac{2(z+1)}{r_h}-\frac{mz}{r_h^{z+3}}\right)\quad {\rm and}\quad s=\frac{r_h^2}{4G_4 l^2},
\ee
where we have denoted the location of the outer event horizon to be $r=r_h$, i.e. $f(r_h)=0$.

\section{The optical conductivity}
\label{sect:conduc}
In this section, we will numerically compute the optical conductivity of the non-relativistic quantum field theory dual to the above charged Lifshitz black brane \footnote{ The optical conductivity studied in the present paper is calculated from the current-current 2-point correlator of the bulk perturbed $U(1)$ gauge field $A_2$ in the linear perturbation limit, i.e., the bulk effective action is expanded up to the second order which corresponds to the tree level approximation from the boundary quantum field theory side. Note that the optical conductivity of the disordered solids is related to the hopping of electrons or ions, thus it is more appropriate to consider the charge transport of the fermions, namely, to compute the optical conductivity from the bulk charged fermions. However, such optical conductivity is from the one-loop bulk effective action contribution, see \cite{Iqbal:2011ae,Faulkner:2013bna} for related studies in AdS spacetime. We expect that the optical conductivities obtained by these two methods reveal different aspects of the boundary condensed matter system and we will further study their relationship in another work.}. We will show that the holographic optical conductivity is qualitatively similar to those in various of disordered solids in condensed matter systems. In order to calculate the electric conductivity $\sigma$, we will turn on the $x-$direction perturbation of the gauge field $A_2$, and also the perturbation along $tx$-component of the metric, {\it i.e.}, we will work out the electric conductivity with backreaction.\footnote{From \cite{Tarrio:2011de}, we know that $A_{1t}$ is divergent at the boundary $r\to\infty$, it plays a key role in supporting the geometry of the Lifshitz spacetime instead of contributing to the free charge of the electromagnetic field; While $A_{2t}$ is the real free electromagnetic field. Besides, our numeric results also show that the perturbation of $A_1$, say $a_{1x}$, is divergent at the boundary. Thus only the fluctuations of $A_{2}$, namely, $a_{2x}$ is the genuine electromagnetic perturbations, which will contribute to the electric conductivities. Therefore, we only need to turn on the perturbations $a_{2x}$ and $h_{tx}$, while turning off the perturbation $a_{1x}$. In the following we will refer to the perturbations $a_{2x}$ as $a_x$.}  Specifically, we will set the perturbation of gauge fields as $\delta A_{x}(t,r)=a_{x}(r)e^{-i\omega t}$, as well as the perturbations of the metric along $tx$-direction to be $\delta g_{tx}(t,r)=h_{tx}(r)e^{-i\omega t}$. \footnote{ The fluctuation $a_{2x}$ is the transverse channel and $h_{tx}$ is the shear channel, they are both vector modes and their EoMs can be obtained by applying the combined variation (diffeomorphism + $U(1)$ gauge transformation) to the bulk Maxwell equation, which only results in two independent EoMs for $a_{2x}$ and $h_{tx}$ as eq.(\ref{eomhtx}) and eq.(\ref{eoma2x}). While other linear perturbations are decoupled with the vector mode and will not affect the calculation on the optical conductivity.} It turns out that the coupled linear EoMs for the perturbations are,
 \be\label{eomhtx}
 h_{tx}'-\frac2r h_{tx}+\frac{qa_{x}}{r^{3-z}}&=&0,\\
\label{eoma2x} a''_{x}+\left(\frac{f'}{f}+\frac{z+1}{r}+\lambda_2\frac{d\phi}{dr}\right)a'_{x}+\left(\frac{\om^2l^{2z+2}}{f^2r^{2z+2}}-\frac{q^2\mu^{-\sqrt{2\frac{z-1}{d-1}}}l^{2z}}{f r^{4+2z}}\right)a_{x}&=&0.
  \ee
  For $r\to\infty$, Eq.\eqref{eoma2x} will become
 \be\label{expa2x}
 a''_{x}+\frac{d+3z-4}{r}a'_{x}+\frac{\om^2l^{2z+2}}{r^{2z+2}}a_{x}&=&0.
 \ee
 The general solution for Eq.\eqref{expa2x} is
 \be
 a_{x}= \frac{\mathfrak{C}_1}{r^{(d+3z-5)/2}} J_{\frac{(d+3z-5)}{2z}}\left(\frac{\om l^{z+1}}{z r^z}\right)+ \frac{\mathfrak{C}_2}{r^{(d+3z-5)/2}} J_{-\frac{(d+3z-5)}{2z}}\left(\frac{\om l^{z+1}}{z r^z}\right)
 \ee
in which, $J_{\pm\al}(\beta)$ is the Bessel function of the first kind, while $\mathfrak{C}_1$ and $\mathfrak{C}_2$ are the constant coefficients. The explicit asymptotic behavior of $a_{x}$ near the infinite boundary for $1\leq z\leq 2$ can be found as (We have set $l=1$.)
\be
\label{expansion}
a_{x}({r\rightarrow\infty})\sim\left\{
\begin{array}{ll}
C_1+\frac{C_2}{r^{3z-2}}, & \quad  1\leq z<2, \vspace{.2cm}\\
C_1+\frac{C_1\omega^2\log(r)}{4r^{4}}+\frac{C_2}{r^{4}}, & \quad  z=2.
\end{array}
\right.
\ee
 in which $C_1$ and $C_2$ might be functions of $\om$ depending on which $z$ we choose. Explicitly, $C_1$ and $C_2$ are constants for $z=1$; $C_1$ is a constant while $C_2=d_2 \omega^{5/3}$ for $z=3/2$ where $d_2$ is a constant; $C_1$ is a constant while  $C_2=\omega^2\left(d_2-\frac{1}{8}C_1\log(\omega)\right)$ for $z=2$. These coefficients $C_1$ and $C_2$ (or $d_2$) should be determined by integrating the Eq.\eqref{eoma2x} from the horizon to the boundary via numerical methods.  According to the dictionary of gauge/gravity duality, $C_1$ represents the source while $C_2$ represents the vacuum expectation value of the current operator $\mathfrak{J}_x$ dual to $a_{x}$. In addition, the asymptotic behavior of $h_{tx}$ near the infinity boundary is,
 \be h_{tx}\sim r^2 h_{tx}^{(0)}+\frac{h_{tx}^{(1)}}{r^{2-z}}+\cdots,\ee
where, $h_{tx}^{(1)}=C_1 q/(4-z)$ in which $C_1$ is the source term of the expansions in $a_{x}$, see Eq.\eqref{expansion}.

\subsection{Quadratic renormalized on-shell action}
In \cite{Sun:2013wpa}, we have derived the on-shell renormalized quadratic action from which the conductivity can be obtained. Therefore, we will just show the results in the following without detailed calculations. The total renormalized quadratic on-shell action for the perturbations $a_{x}$ and $h_{tx}$ is,
\be
I^{(2)}_{\text{total}}
=\int d^dx ~\left(C_1C_2(3z-2)-(4-z)h_{tx}^{(0)}h_{tx}^{(1)}-{m}(h_{tx}^{(0)})^2r^{2-2z}\right),\quad
\ee
for $1\leq z<2$. Or,
\be
I^{(2)}_{\text{total}}
&=&\int d^dx ~\left(C_1C_2(3z-2)-\frac{C_1^2\om^2}{3z-2} -(4-z)h_{tx}^{(0)}h_{tx}^{(1)}-{m}(h_{tx}^{(0)})^2r^{2-2z}\right)\nno\\
&=&I^{(1)}_{\text{total}}-\int d^dx \frac{C_1^2\om^2}{3z-2},
\ee
when $z=2$. Therefore, we can readily calculate the optical conductivity $\sigma(\om)$ as the following:
\be\label{conduc}
 \sigma(\omega)&=&\left\{
  \begin{array}{ll}
    \frac{C_2(3z-2)}{i\om C_1}, &  \quad {1\leq z<2;}  \vspace{.3cm}\\
        \frac{C_2(3z-2)}{i\om C_1}-\frac{2\om}{i(3z-2)}, &  \quad {z=2.}
  \end{array}
\right.
\ee

\subsection{Numerical results of conductivity}
In the numerical calculations, we have adopted the usual ingoing boundary conditions near the horizon and scaled
$l=1, r_h=1$, and $\mu=1$.

\begin{figure}[h]
  \hspace{4mm} \includegraphics[scale=0.652]{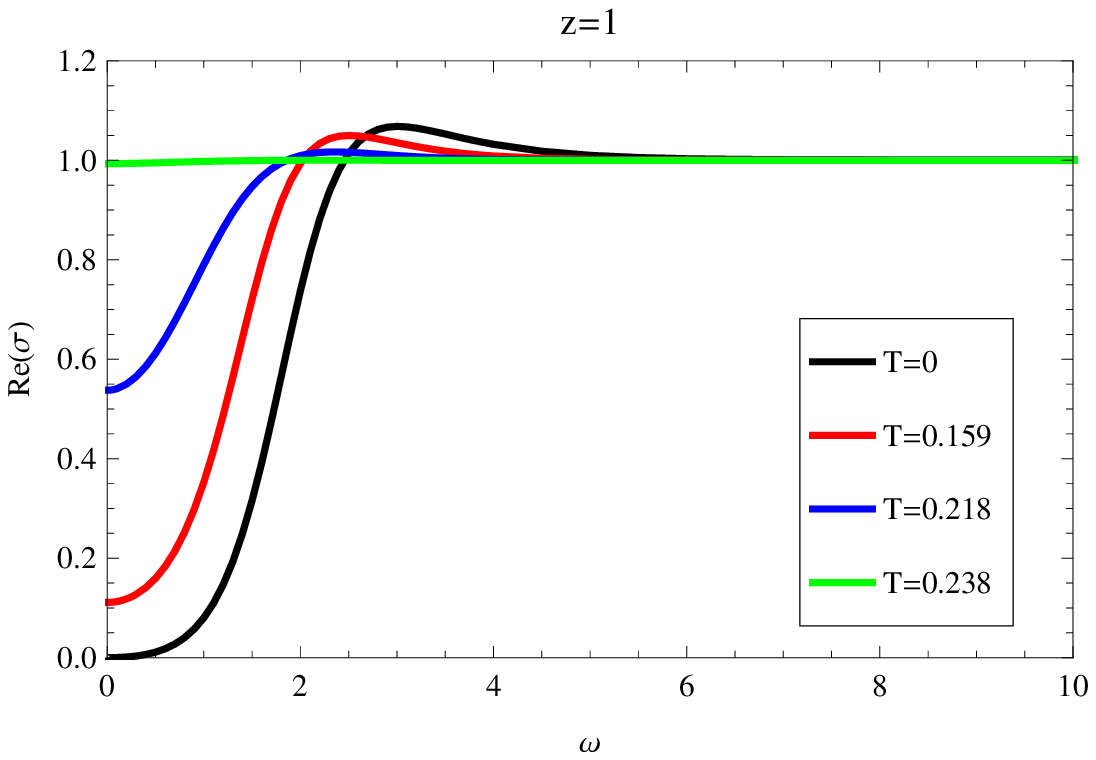}
   \hspace{-1cm}\includegraphics[scale=0.672]{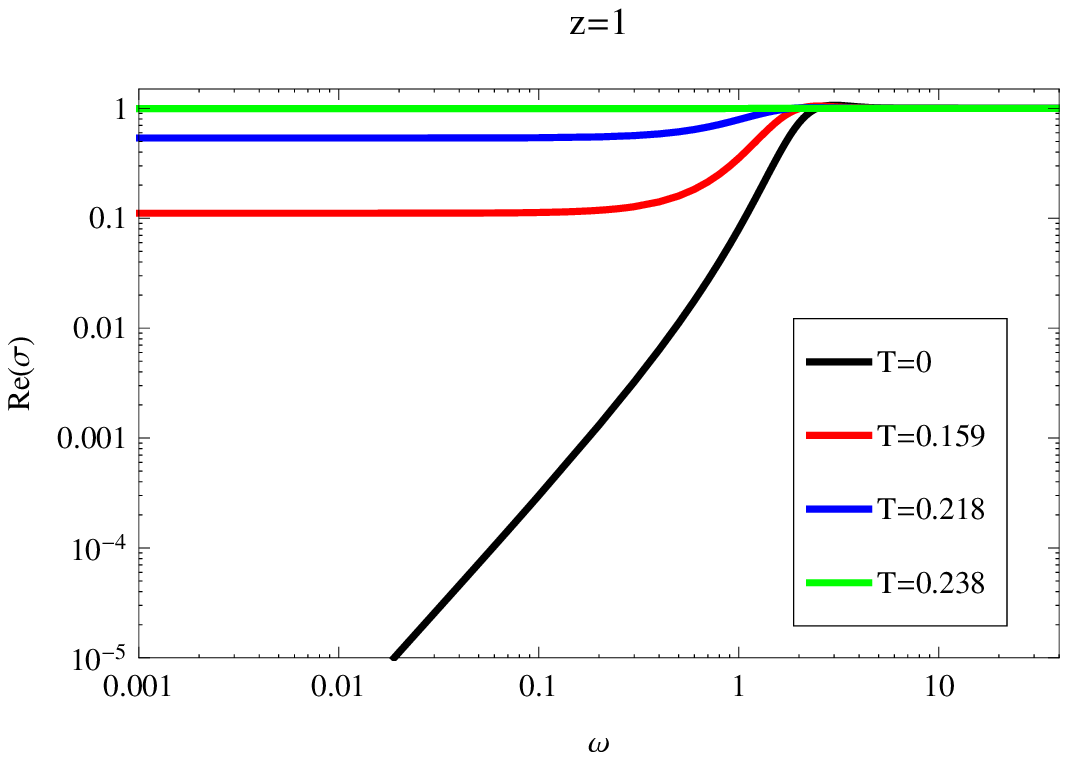}
  \includegraphics[scale=0.55]{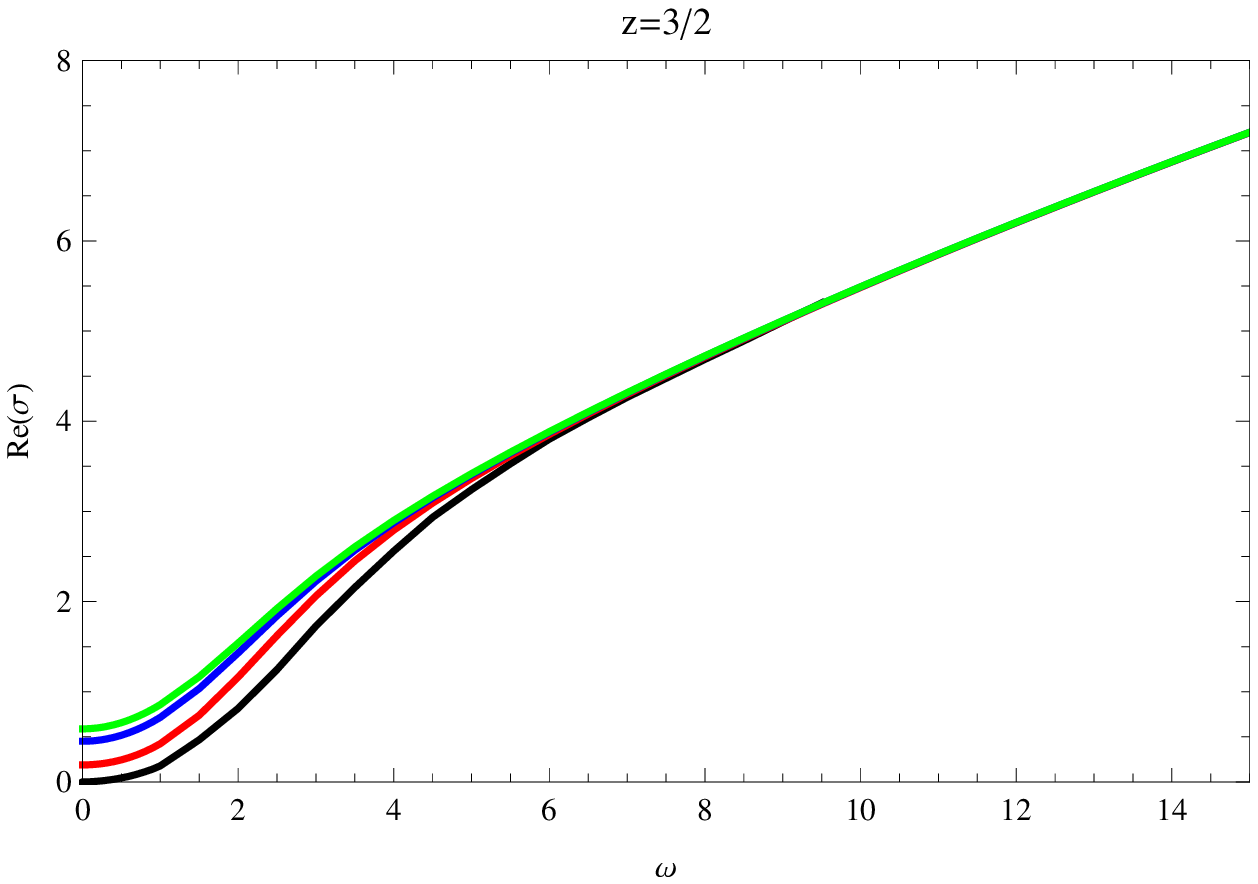}
   \hspace{.2cm}\includegraphics[scale=0.58]{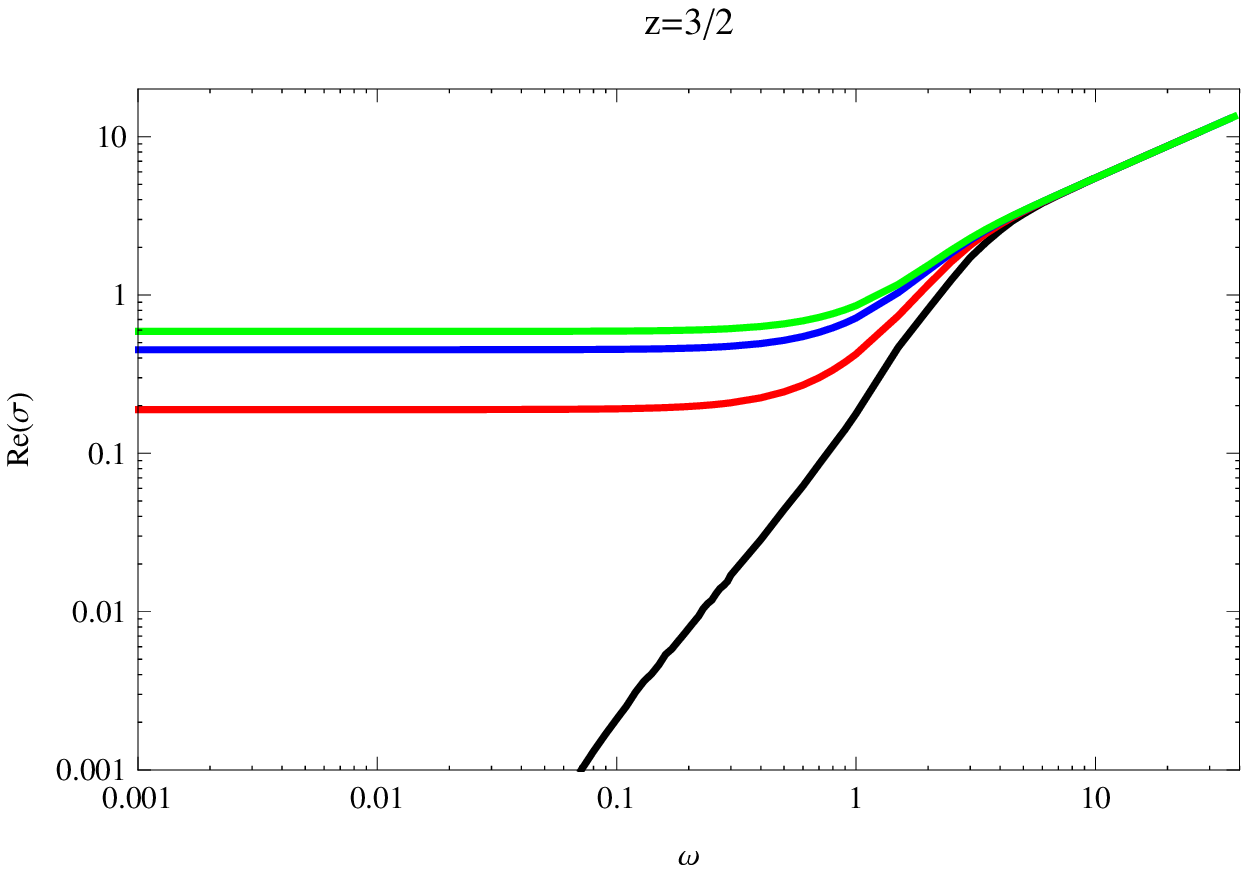}
   \includegraphics[scale=0.55]{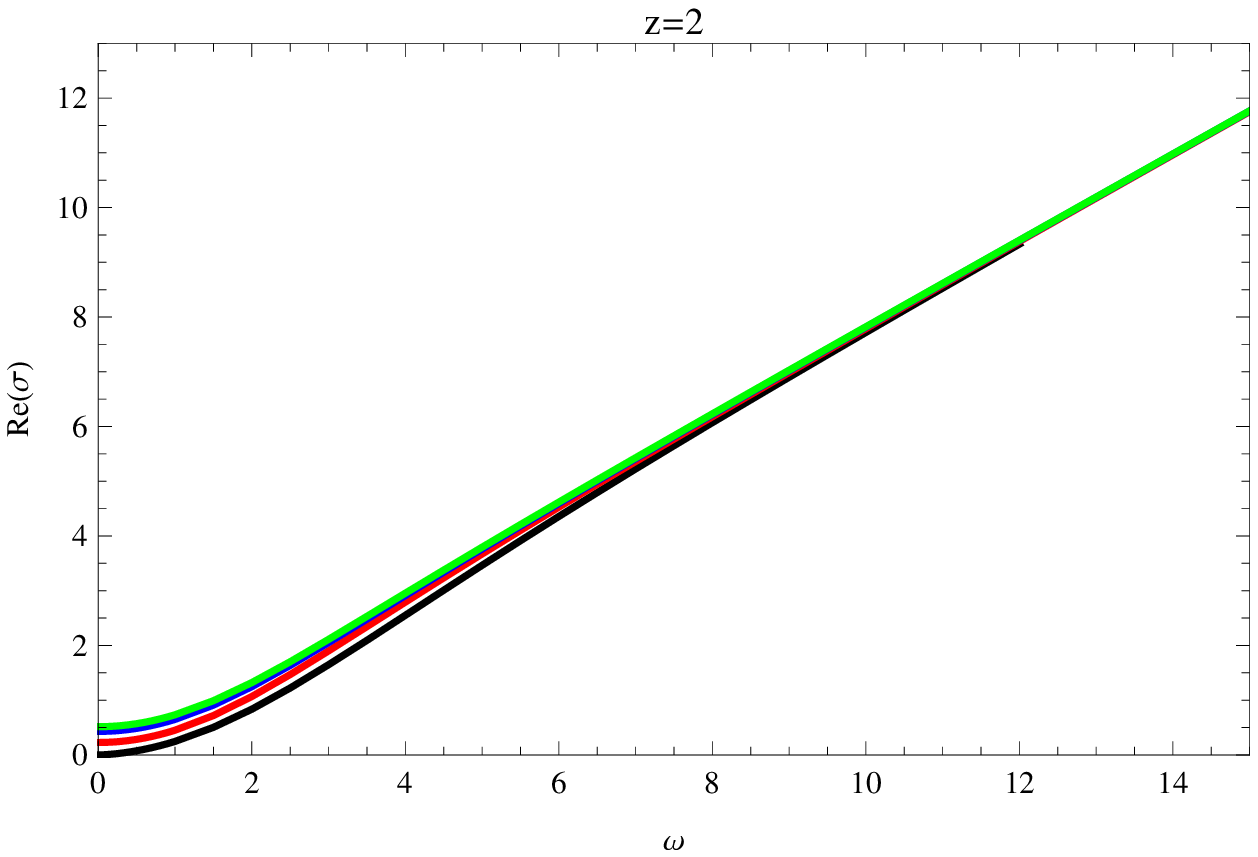}
    \hspace{.2cm}\includegraphics[scale=0.58]{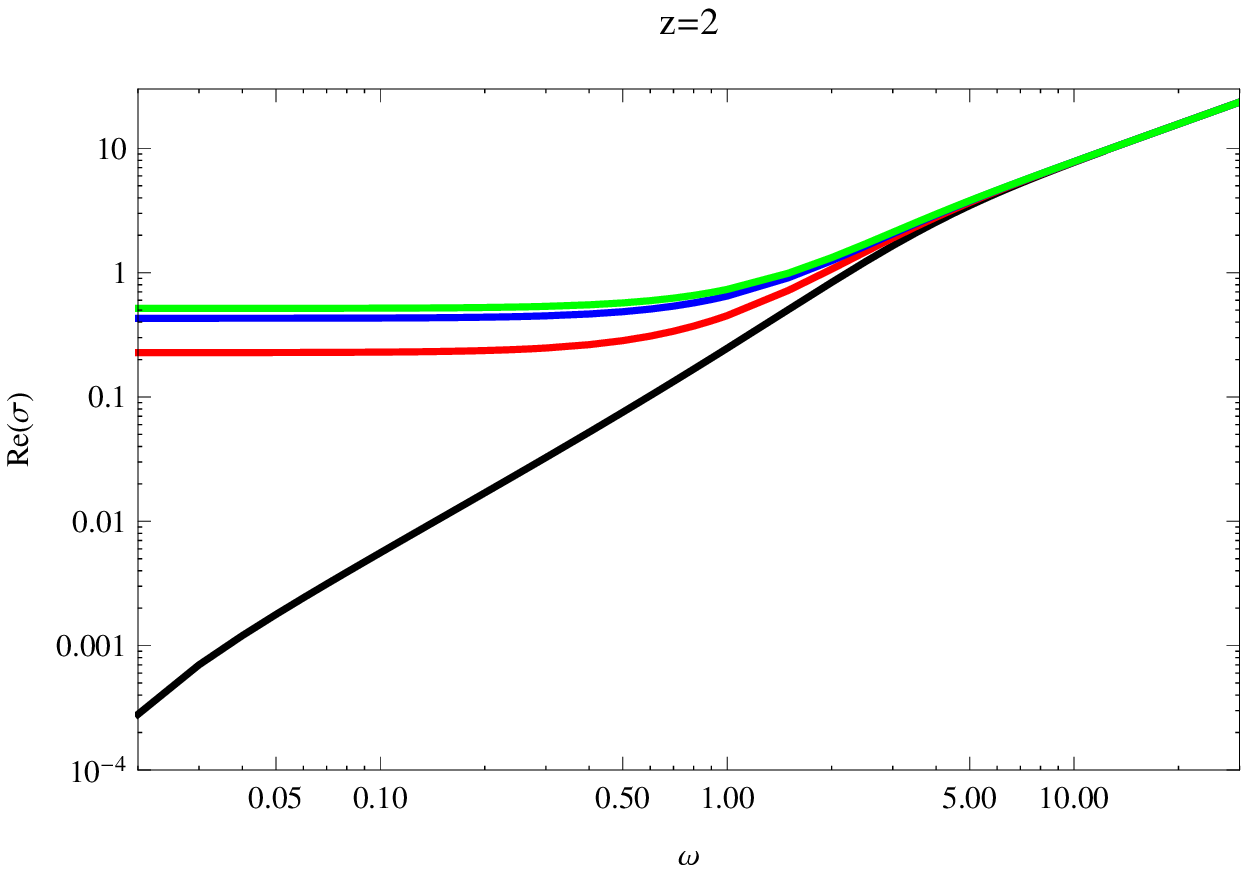}
  \caption{\label{d3}(Left) The real part of conductivity versus the frequency for different temperatures $T$ and various $z$; (Right) The log-log plot of the real part  of the conductivity versus frequency for different temperatures $T$ and various $z$.}
\end{figure}
\subsubsection{Figure \ref{d3}}
In Fig.\ref{d3}, we show the real part of conductivity Re($\si$) versus the frequency $\om$, for different temperatures $T$ and various $z$ in the left panel. In order to compare the results to those in condensed matter physics, we also draw the log-log plot for the conductivity versus the frequency in the right panel.

$\bullet$ {\bf High frequency regime:}~ For large frequencies, the asymptotic behavior of $\sigma$ can be obtained from the expansions in Eq.\eqref{expansion} and the formula in Eq.\eqref{conduc}, it is
\be\label{largeom}
 \sigma(\omega\gg1)\sim\left\{
  \begin{array}{cl}
    \omega^0, & {z=1;} \\
    \omega^{2/3}, & {z=\frac 3 2;} \\
    \omega(a+\log(\om)), & {z=2.}
  \end{array}
\right.
\ee
where, $a$ is a certain constant independent of $\omega$. We can clearly see that these large frequency scaling behaviors of the conductivity are consistent with those in the left panel of Fig.\ref{d3}.

For high frequency, the real part of the conductivity will tend to a constant when  $z=1$, which is similar to the results in studying holographic superconductors, e.g. \cite{Hartnoll:2009sz}. But the differences come from the cases of $z=\frac 3 2$ and $z=2$, in which the Re($\sigma$) will depend on $\omega$ according to Eq.\eqref{largeom}. This is an interesting and new phenomenon from the viewpoint of the gauge/gravity duality, which has not been observed in the previous literatures as far as we know. Actually, people always argue that the electric conductivity in $(2+1)$-dimensional field theory will tend to a constant in the high frequency region, because conductivity has a vanishing scaling dimension in this spacetime dimension \cite{Hartnoll:2009ns}. However, in condensed matter physics, the electric conductivity will not always tend to a constant even in $(2+1)$-dimensional field theory. For instance, in \cite{Dyre:2000zz} the author studied the optical conductivity for various disordered solids in $(2+1)$ dimensions both from the experiments and model building, and showed that there does exist a nontrivial scaling behavior of the electric conductivity in the large frequency regime, which is independent of details of the disordered solids.

In order to compare our holographic results to the results in condensed matter physics, we also plot the log-log figure of the conductivity versus the frequency on the right panel in Fig.\ref{d3}. It can be found that the behavior of the conductivity is very similar to those in \cite{Dyre:2000zz}, at least qualitatively. We need to point out that although the coincidence of the holographic conductivity with those in disordered solids are very surprising, the underlying detailed correspondence between these two aspects are still unclear so far. Unlike previous attempts on studying the impurities and disordered systems via gauge/gravity duality in which interactions randomly distributed spatially are added in the action \cite{Hartnoll:2008hs,Fujita:2008rs,Ryu:2011vq,Hashimoto:2012pb}, there are no such interaction terms introduced in our model instead of two spatial uniformly distributed $U(1)$ gauge fields coupled with the dilaton in Eq.(\ref{EMd1}). However, we find that the peculiar frequency dependence of the holographic optical conductivities in Eq.(\ref{largeom}) stem from the coupling of the dilaton with the second Maxwell field in Eq.\eqref{EMd1}, {\it i.e.} the $e^{\lambda_2\phi}F_2^2$ term. Note that although the second $U(1)$ gauge field does not interact with the first one directly, they indeed interact with each other indirectly via the dilaton field. In this sense, we regard the fluctuation of the second gauge field $a_x$ as the impurity field, it may relate to the disorder parameter of disordered solids in the extreme disorder limit when local randomly varying mobilities of charge carriers cover many orders of typical length scale of the condensed matter system, which results in the homogeneously distributed local optical conductivity \cite{Dyre:2000zz}.

In order to see more clearly how we obtain this particular feature in the optical conductivity, we compare our calculations with those performed in the Lifshitz black brane which is a solution of the EMD theory with only one $U(1)$ gauge field (it has a similar role as $A_1$ in our case) \cite{Pang:2009wa}. Rather than two gauge fields case, the single $U(1)$ gauge field can only supports the geometry of a neutral Lifshtiz spacetime. To study the conductivities, additional probe gauge field needs to be added into the background, it is shown that this kind of probe gauge field actually had the same $r$ dependence as $A_2$ in our paper. However, the key difference was that the probe $U(1)$ gauge field added in \cite{Pang:2009wa} is minimally coupled, namely, it does not couple with the background dilaton. However, we have a natural interaction between the dilaton and the gauge field $A_2$ in the action and consequently, the effect of the dilaton $\phi$ will enter into the coefficient of $a'_x$, {\it i.e.}, through the $(\lambda_2 d\phi/dr)$ term in Eq.\eqref{eoma2x}.\footnote{Note that the dilaton field has a logarithmic behavior in $r$, {\it i.e.}, $\phi\sim\log(r)$ which is important to render $d\phi/dr\sim 1/r$. And then this $1/r$ behavior will enter into the coefficients of $a'_x$ at the boundary $r\to\infty$.} Besides, the rest terms in Eq.\eqref{eoma2x} are the same as those in \cite{Pang:2009wa} when approaching the boundary $r\rightarrow \infty$. Therefore, this dilaton contribution to the coefficients of $a'_x$ is very crucial. It will modify the expansions of $a_x$ near the boundary, and finally make the conductivity varies as well. \footnote{\label{f2} Turning on the backreaction in the $tx$ component will only affect the conductivity in the low frequency regime. Because the metric fluctuations will only enter into the coefficients of $a_x$, and it has a higher order of $1/r$ than the term contains $\om^2$. In our paper, it is $1/r^{2z+4}$ which will decay more rapidly than the term $1/r^{2z+2}$ which contains $\om^2$ when $r\to\infty$, please see Eq.\eqref{eoma2x}. Therefore, the backreaction will have a minor modification on the large frequency behavior of the conductivity. This can also be understood physically, because for large frequency the frequency energy will be dominant than other ingredients, such as the temperature, chemical potential, etc. This is also the origin of the universality that the high frequency behavior of the conductivity is robust against the temperature.} Physically, we speculate that the interactions between the dilaton and the Maxwell fields will give extra dimensional scale into the boundary field system and consequently render the conductivity not to tend to a constant at high frequency. Although there are a lot of works studying the conductivity in dilaton gravity from the gauge/gravity duality before, such as \cite{Goldstein:2009cv,Charmousis:2010zz,Liu:2010ka,Salvio:2012at,Salvio:2013jia}, they mainly focus on the relativistic case, {\it i.e.}, $z=1$. Our present work is a further step towards studying the AC optical conductivity of the non-relativistic quantum field theories dual to the dilatonic-like Lifshitz spacetime, in the hope of describing the real condensed matter systems in the laboratory conditions. \footnote{In \cite{Alishahiha:2012qu}, the authors worked in a dilatonic-like Lifshitz spacetime with hyperscaling violation factors. However, they studied the conductivity for a special case $\theta=d-1$, where $\theta$ is a hyperscaling violation exponents, while $d$ is the spacial dimension of the boundary. In order to compare their spacetime background to ours, one should impose $\theta\equiv0$, {\it i.e.}, $d=1$. Therefore, they studied the conductivity for a $(1+1)$-dimensional field theory, if compared to our backgrounds.}

We also find that the high frequency behavior of the conductivities have the same slope whatever the temperature is. This kind of universality of the optical conductivity was also similar to those analyzed in \cite{Dyre:2000zz}. This property can be understood from what we have mentioned in the footnote \ref{f2} that for large frequency, the frequency energy will dominant, therefore, the temperature effect will be minor.

$\bullet$ {\bf Low frequency regime:}~In the low frequency regime, we set the smallest frequency be $10^{-5}$ in the numerical calculations, because when $\om=0$ there will be some numerical instability. In the following, we will regard the DC conductivity to be the conductivity at $\om=10^{-5}$ in practice, rather than the conductivity at $\om=0$ in theory. We think that this will not lead to any misunderstanding. First, we can see that the DC conductivity will increase as the temperature increases. Particularly, when the temperature is zero the DC conductivity will vanish for any $z$. Physically this means the conducting electrons are `frozen' at $T=0$. Second, from the right panel of Fig.\ref{d3}, we find that when the temperature is a little bigger than zero, the slope of the conductivity in low frequency is very flat. The exception occurs when $T=0$, at which the slope is much deeper than those at a finite $T$. The above low frequency behaviors of the holographic optical conductivity are also consistent with those in \cite{Dyre:2000zz}.

\begin{figure}[h]
 \hspace{4mm}  \includegraphics[scale=0.65]{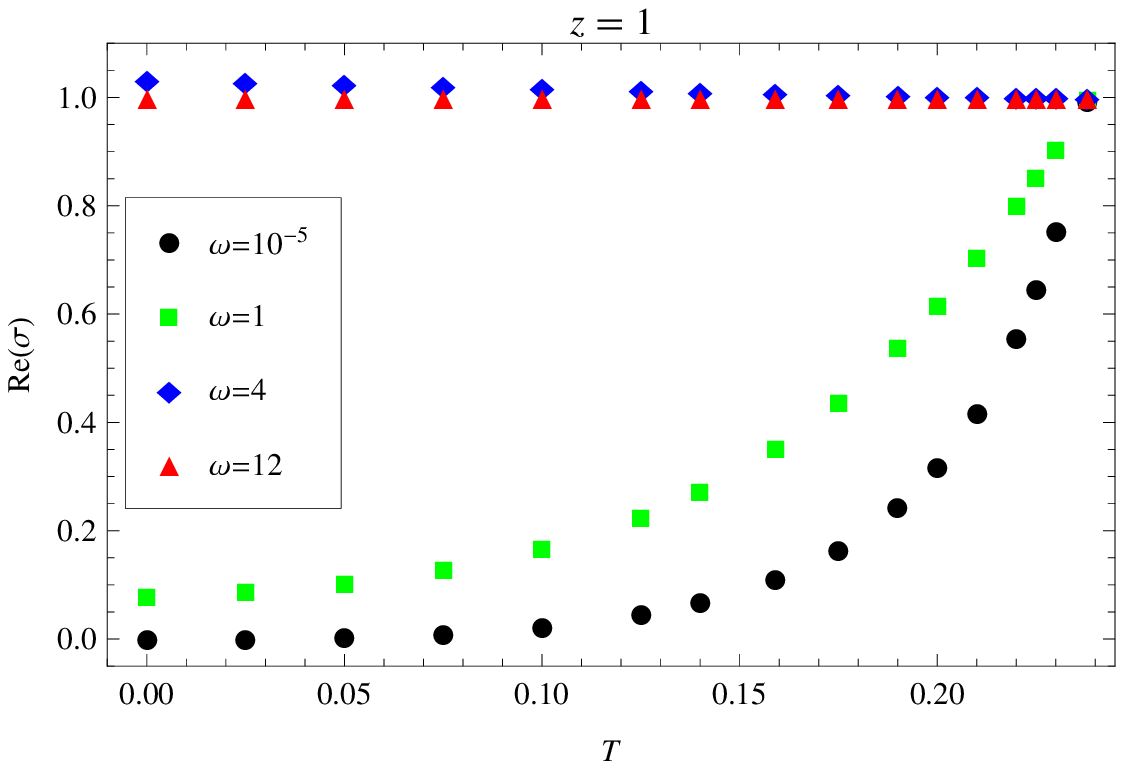}
   \hspace{-.8cm}\includegraphics[scale=0.65]{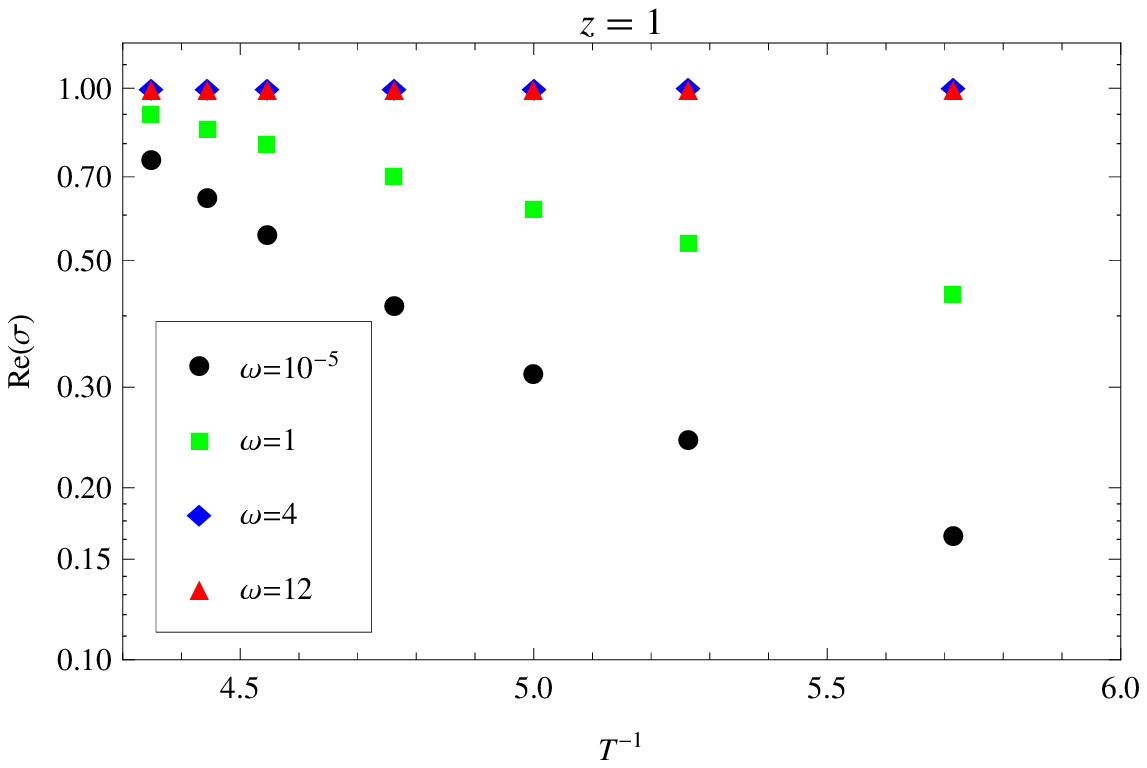}
  \includegraphics[scale=0.55]{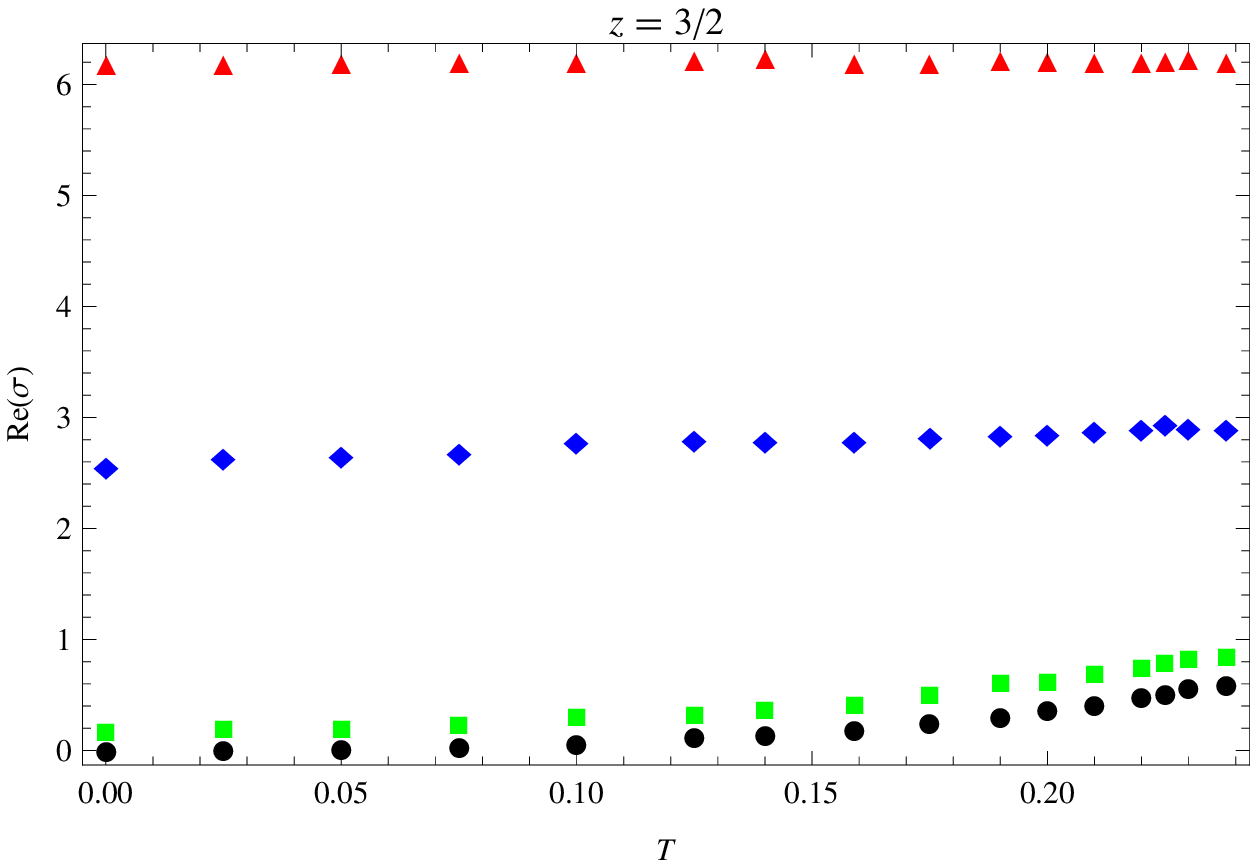}
  \hspace{.2cm}\includegraphics[scale=0.58]{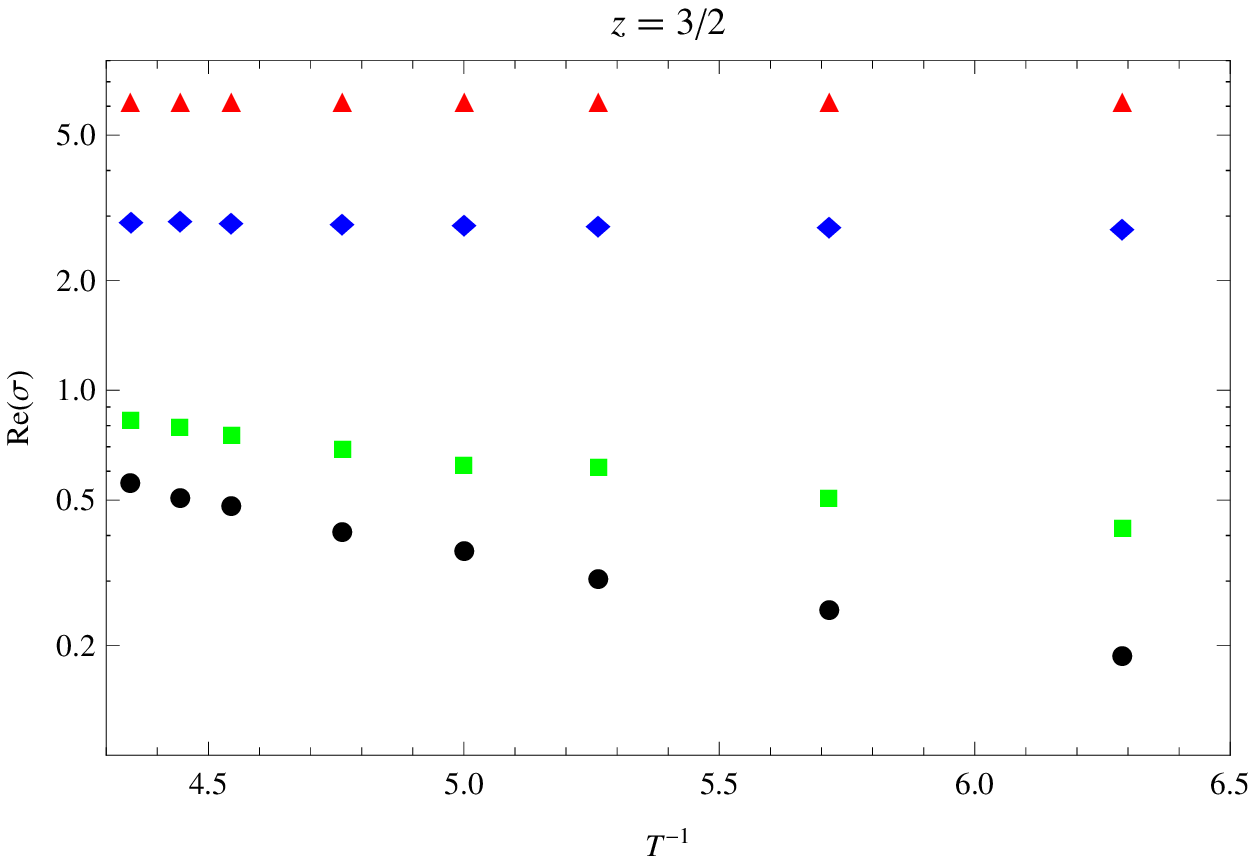}
   \includegraphics[scale=0.55]{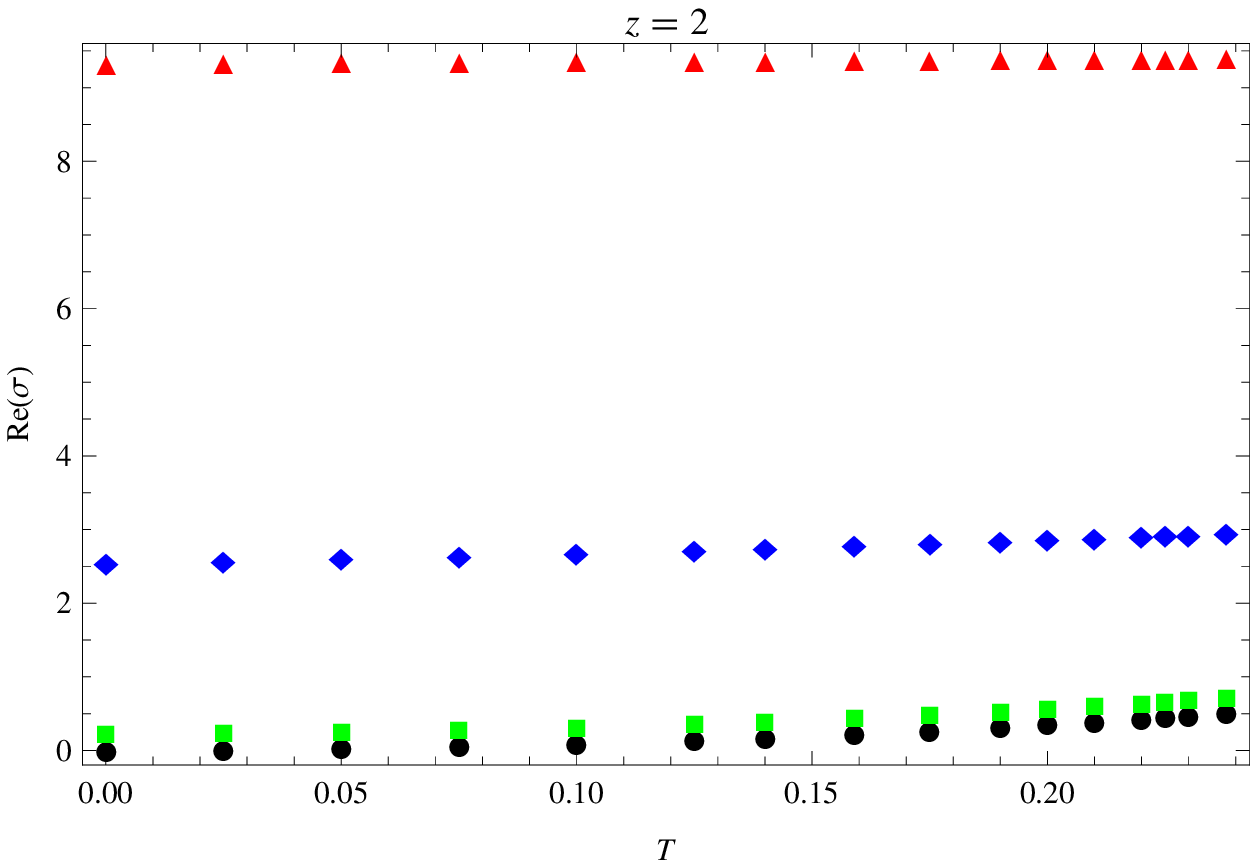}
   \hspace{.2cm}\includegraphics[scale=0.58]{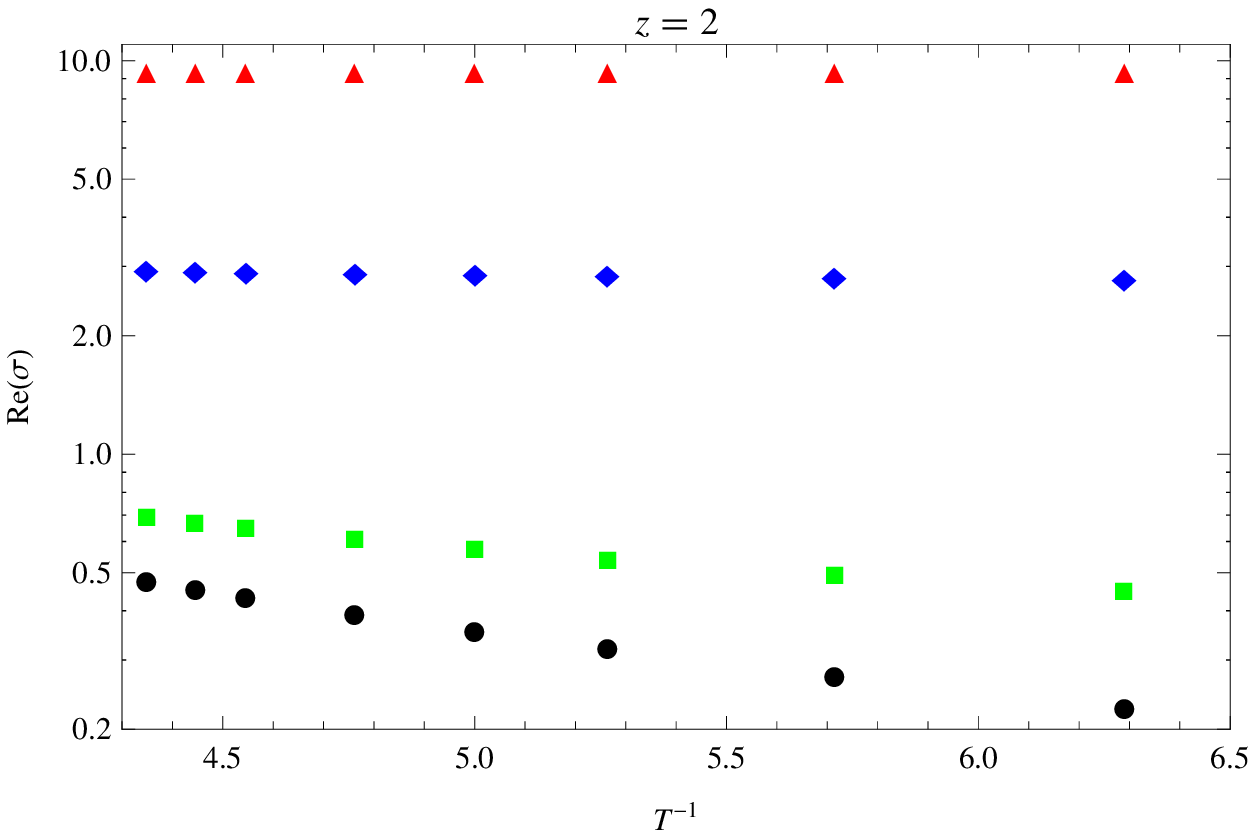}
  \caption{\label{fixom} (Left) The real part of conductivity versus temperatures for different frequency $\om$ and various $z$; (Right) The log plot of the conductivity versus the reciprocal of temperatures for different frequency $\om$ and various $z$. }
\end{figure}

\subsubsection{Figure \ref{fixom}}
 In the left panel of Fig.\ref{fixom}, we plot the real part of the conductivity versus the temperature under different frequency and various $z$. We find that for a fixed temperature, when frequency increases the conductivity will increase as well. And for a fixed low frequency (say $\om=10^{-5}$ or $\om=1$), the conductivity will grow as the temperature grows; However, for a fixed high frequency (say $\om=12$), the conductivity will always be flat with respect to temperature. This reflects the universality we have mentioned above. The conductivity for $\om=4$ is in between with the low and high frequency, it will grow more moderate according to the temperature. The exception is for $z=1$, because in this case the conductivity for $\om=4$ is in the bump part in the upper-left plot for Fig.\ref{d3}. The above analysis is consistent with the Fig.\ref{d3}, so we will not discuss them at length.

 In order to compare the above results to the condensed matter literature more apparently, such as \cite{Psarras:2006}, we also draw the log plot of the conductivity versus the reciprocal of the temperature $T^{-1}$ in the right panel of Fig.\ref{fixom}. The range of $T^{-1}$ is roughly $4.3\sim6.5$, therefore, the temperature $T$ is about $0.15\sim0.23$ which is in the higher temperature regime in the left panel of Fig.\ref{fixom}. We can readily find that the $\log(Re(\si))$ is linearly proportional to $T^{-1}$ in this temperature regime. This linear behavior is consistent with the results in \cite{Psarras:2006} qualitatively. In paper \cite{Psarras:2006}, the author studied the conductivities in a hopping model which was used to investigate the disordered solids. Therefore, from Fig.\ref{fixom} we find that the behavior of the conductivity is consistent with that in disordered solids in experiment as well. We think that these consistencies should have some deep relations between the dilatonic-like Lifshitz spacetime and the disordered solids, but the more clear physical picture needs to be further explored.

 \begin{figure}[h]
   \includegraphics[scale=0.8]{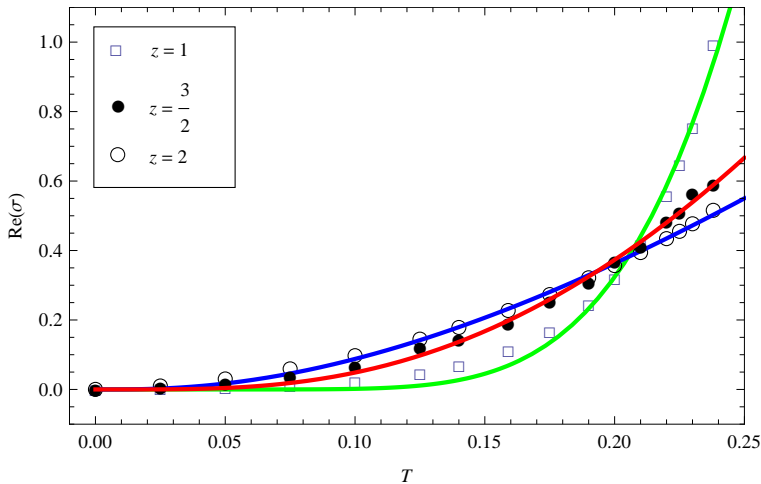}
   \caption{\label{fitting} Fitting of the DC conductivity versus the temperature for various $z$.}
\end{figure}
\subsubsection{Figure \ref{fitting}}

 In the right panel of Fig.\ref{fixom}, we find a linear relation between $\log(Re(\si))$ and the reciprocal of the temperature $T^{-1}$ in a finite range of the temperature.  However, for the whole range of the temperature we considered in this paper, this linear relation will not hold anymore. It was suggested in \cite{Psarras:2006} that the DC conductivity could be approximated by the following relation
 \be\label{fit} \sigma(T)=\sigma_0 \exp\left[-\frac{T_0}{T^\gamma}\right],\ee
 where, $\sigma_0$ represents the conductivity when $T\to\infty$, $T_0$ is a parameter which can characterize the disorder, while $\gamma$ is related to the dimension of the field theory, so in our case $\gamma=1/3$ if we admit the above relation Eq.\eqref{fit}. But in our paper, we will regard $\ga$ as an undetermined parameter which will be set by the data, and then compare the discrepancy between the $\ga$ we deduce with that $\ga=1/3$.

 In Fig.\ref{fitting}, the squares, dots and circles are the data of DC conductivity from Fig.\ref{fixom}, while the solid lines are the fitting curves of the data by virtue of the relation Eq.\eqref{fit}. The parameters we have fitted are shown in Table \ref{tab}.
\begin{table}[h]
\begin{tabular}{|c|c|c|c|}
  \hline
    & $\sigma_0$ & $T_0$ & $\gamma$ \\
  \hline
  $z=1$ & 205755.251 & 6.194 & 0.477 \\
  $z=3/2$ & 9306.565 & 6.611 & 0.264 \\
  $z=2$ & 3812.844 & 6.649 & 0.205 \\
  \hline
  \end{tabular}
  \caption{\label{tab} The fitted values of the parameters in Eq.\eqref{fit} for various $z$.}
\end{table}
We find that $\ga$ will decrease according to the growth of $z$, rather than $\gamma=1/3$ in \cite{Psarras:2006}, however, we can see that $\ga$ we fitted does not departure from $\ga=1/3$ very much. If our holographic model can genuinely mimic the disordered solids, the dynamical critical exponents of the disordered solids must lie between $1<z<3/2$. $T_0$ for various $z$ are around $6$, which does not also change very much. $\sigma_0$ will decrease as $z$ increases, which indicates that the dual field theory for $z=1$ will have the largest conductivity when $T\to\infty$. But as we have discussed above, the deep relations between our holographic model and the disordered solids are still unclear, so the real physical meanings of the parameters $\ga, T_0$ and $\si_0$ are still vague in the gravity side.

\section{Conclusions and Discussions}
\label{sect:conclusion}
In this paper, we calculated the optical conductivity in the background of a 4-dimensional charged Lifshitz black brane with two independent $U(1)$ gauge fields in the EMD theory. When $z=1$, we found that the behavior of the conductivity was similar to previous studies in the AdS/Condensed matter literatures. However, for $z>1$, we observed a peculiar behavior of the optical conductivity which would not tend to a constant in the high frequency regime even in $(2+1)$-dimensional field theory. Strikingly, we found that the (non-)power law scaling behavior of the AC conductivity was consistent with the experiments in various disordered solids at least qualitatively. Furthermore, this kind of scaling behavior showed a universality for a fixed $z$ whatever the temperature was. In the low frequency regime, the conductivity would decrease if the temperature decreased. In addition, we also found a linear relation between the logarithmic of the conductivity versus the reciprocal of the temperature in certain regimes of the temperature. This linear relation was also qualitatively consistent with the experiments in disordered solids. Even though the holographic optical conductivity we obtained performed some special features analogous to those in the disordered solids, we have to point out there was no apparent disorder parameter in our model, namely, there was neither spatial inhomogeneity in the background spacetime, nor interaction terms randomly distributed on the spatial coordinates.\footnote{ Actually in this probe limit, the imaginary parts of the conductivity will diverge at low frequency, this will force the real parts of the conductivity be a delta function at zero frequency due to the Kramers-Kronig relation. Physically, it is due to the translational symmetry in our model because apparently we did not introduce any spatial  inhomogeneity. This delta function in the real parts of the conductivity or the pole in the imaginary parts of the conductivity is a shortcoming of the probe limit for the holographic study of the conductivity transport coefficients without superconductivity. A possible way to avoid this situation is maybe one can introduce a spatial dependent lattice structure into the system, just like the paper \cite{Horowitz:2013jaa} did. However, this is beyond the scope of our current work, we will consider this in our future work.} However, as we have explained in the introduction, physically, the fluctuation of the second $U(1)$ gauge field could be interpreted as the impurity field, which interacted with the first $U(1)$ gauge field indirectly through the dilaton. While the resulting homogeneous optical conductivity may relate to the extreme disorder limit of disordered solids in which local randomly varying mobilities of charge carriers could cover many orders of typical length scale of the condensed matter system. Technically, the peculiar scaling behavior of the optical conductivity actually stemmed from the couplings of the dilaton to the Maxwell fields, which lead to an unusual expansions for the perturbation of gauge fields and thus altered the conductivity. Thus we would like to say that we observed the phenomenon similar to the optical conductivities in disordered solids via the gauge/gravity duality, instead of claiming that we have found a good holographic model that could genuinely describe the disordered solids. We believe that there should have some deep physics to explain the coincidence but the underlying details and the holographic model require further investigations.

\section*{Acknowledgement}
We would like to thank Jeppe C. Dyre, Sean A. Hartnoll and Stefan Vandoren for helpful communications. J.R.S. was supported by the National Science Foundation of China under Grants No. 11147190 and No. 11205058; S.Y.W. was supported by the National Science Council (NSC 101-2811-M-009-015) and National Center for Theoretical Science, Taiwan; H.Q.Z. was supported by a Marie Curie International Reintegration Grant PIRG07-GA-2010-268172.


\end{document}